# In-situ three-dimensional strain engineering of solid-state quantum emitters in photonic structures towards scalable quantum networks

Yan Chen,[1,2,*] Xueshi Li,[2,†] Shunfa Liu,[3] Jiawei Yang,[3] Yuming Wei,[4] Kaili Xiong,[2] Yangpeng Wang,[3] Jiawei Wang,[5] Pingxing Chen,[2,6] Xiao Li,[1] Chaofan Zhang,[1] Ying Yu,[3,6] Tian Jiang,[2,‡] and Jin Liu[3,§]

[1]*College of Advanced Interdisciplinary Studies, National University of Defense Technology, Changsha, 410073, China.*
[2]*Institute for Quantum Science and Technology, College of Science, National University of Defense Technology, Changsha, 410073, China.*
[3]*State Key Laboratory of Optoelectronic Materials and Technologies, School of Physics, School of Electronics and Information Technology, Sun Yat-sen University, Guangzhou, 510275, China.*
[4]*Key Laboratory of Optoelectronic Information and Sensing Technologies of Guangdong Higher Education Institutes, Jinan University, Guangzhou 510632, China*
[5]*School of Electronics and Information Engineering, Harbin Institute of Technology, Shenzhen, 518055, China*
[6]*Hefei National Laboratory, Hefei 230088, China*
(Dated: April 3, 2025)

**Solid-state quantum emitters are pivotal for modern photonic quantum technology, yet their inherent spectral inhomogeneity imposes a critical challenge in pursuing scalable quantum network. Here, we develop a cryogenic-compatible strain-engineering platform based on a polydimethylsiloxane (PDMS) stamp that is not obviously working properly at cryogenic temperature. In-situ three-dimensional (3D) strain control is achieved for quantum dots (QDs) embedded in photonic nanostructures. The compliant PDMS enables independent tuning of emission energy and elimination of fine structure splitting (FSS) of single QDs, as demonstrated by a 7 meV spectral shift with a near-vanishing FSS in circular Bragg resonators and an unprecedented 15 meV tuning range in the micropillar. The PDMS-based 3D strain-engineering platform, compatible with diverse photonic structures at cryogenic temperature, provides a powerful and versatile tool for exploring fundamental strain-related physics and advancing integrated photonic quantum technology.**

Solid-state quantum emitters are essential to the advancement of quantum networks by enabling the deterministic quantum light sources[1–4] and efficient spin-photon interfaces[5]. In the quantum network, efficient indistinguishable sources of non-classical light are required to establish quantum links between remote nodes via two-photon interference[6–9]. Semiconductor quantum dots (QDs) stand out as exceptional candidates due to their high quantum efficiency, photon indistinguishability, and seamless integration with existing semiconductor technology[10]. Nonetheless, naturally grown QDs suffer from low source brightness due to the total internal reflection of the high refractive index semiconductor and spectral inhomogeneity associated to the self-assembly growth. The source brightness can be greatly enhanced by placing single QDs into an engineered photonic environment provided by semiconductor nanostructures, enabling the realization of individual near-optimal single-photon and entangled pairs sources. So far, circular Bragg resonators (CBRs) and micropillars are two representative photonic structures for the generation of high-quality entangled and single photons with embedded QDs[11–16]. However, the natural next step of scaling multiple high-performance semiconductor quantum light sources to a functional quantum network has proven to be extremely challenging because of the solid-state nature of QDs[17, 18], which prevents the efficient spectral overlapping between QDs and high-Q microresonators or between different QDs. While alternative growth methods like droplet etching[19] and site-controlled QDs [20] can reduce inhomogeneity by an order of magnitude compared to Stranski-Krastanov (SK) QDs, their emission wavelengths still exhibit variations exceeding the radiative linewidth, necessitating post-growth tuning for scalable quantum networks. Additionally, entangled photons can only be generated via the biexciton (XX)-exciton (X) cascade in QDs if the fine structure splitting (FSS) is smaller than the linewidth of the intermediate X[21]. Addressing these challenges requires novel post-tuning schemes capable of independently controlling the emission wavelength and FSS, particularly for QDs embedded in photonic structures. To date, the coupling of QDs to the cavity and other QDs has been predominantly based on temperature tuning, the quantum confined Stark effect, and strain tuning[16, 22]. However, temperature tuning is not favored due to a limited tuning range and increased phonon dephasing at elevated temperatures[23, 24]. Stark tuning and strain tuning are not directly compatible with photonic structures, especially for those with very high aspect ratios [25–31].

In this work, we present a universal and high-efficiency technology for generating in-situ 3D strain and reversibly engineering the FSS and emission wavelengths of single QDs embedded in photonic structures, which are essential for connecting disparate quantum nodes via photon interferences. To showcase the power of our tuning method, we apply 3D strain to QDs in CBRs and micropillars as schematically illustrated in Fig. 1. For the CBRs, the relatively low quality(Q)-factor ensures the spectral overlap between the cavity mode and the X and XX lines. In this case, we simultaneously eliminate the FSS and tune their emission wavelengths across 7 meV, for the first time, stepping close to the entanglement swapping with other QD-entangled sources. For generating single photons with wide-range spectral tunability and exploring cavity quantum electrodynamics, we employed 3D strain to tune the X line across multiple high-Q cavity resonances, with an unprecedented range of 15 meV, 20 times larger than the state-of-the-art.

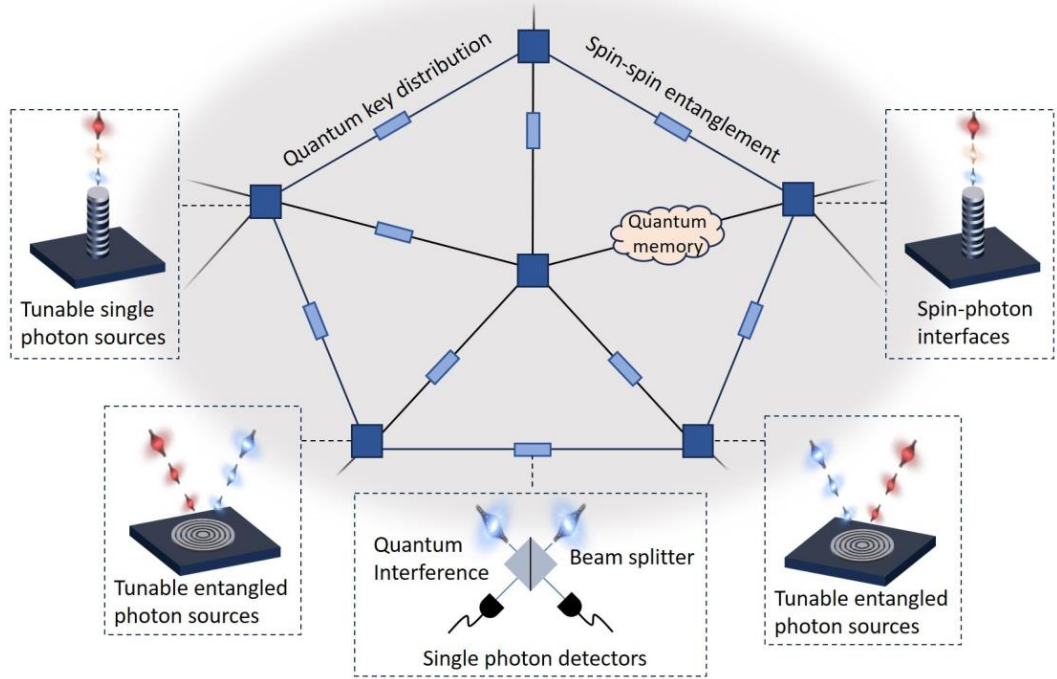

FIG. 1. Future quantum internet with high-performance solid-state quantum light sources as core elements. The multiple single-photon sources in the form of micripillar and entangled pair sources in the form of CBRs have to be tuned to the same wavelength for quantum connections via photon interferences.

The experimental setup involves a cryo-strain apparatus consisting of a multi-axis nano-positioner and a micro-polydimethylsiloxane (PDMS) stamp with a footprint of 50 μm by 50 μm. The stamp is affixed to a highly transparent glass substrate. The glass, nano-positioner, and the QD sample are enclosed within a custom-made titanium box which has a window atop to allow the excitation laser and photoluminescence (PL) signal to pass through. The size of the strain apparatus is around 30mm × 15mm × 60mm, small enough to accommodate into an optical helium bath cryostat in order to avoid temperature problems. The setup is illustrated in Fig. 2a and the fabrication of the cryo-strain apparatus is elaborated in the method section.

A pivotal advantage of our methodology stems from the preserved low Young's modulus of PDMS under cryogenic conditions. When the PDMS stamp is compressed onto the QD sample, it generates a uniformly distributed out-of-plane (OP) strain $S_{zz}$ across the contact interface, which directly modulates the semiconductor bandgap through hydrostatic pressure effects, thereby enabling deterministic tuning of the QD emission energy. Simultaneously, the viscoelastic adhesion at the PDMS-sample interface under compression allows for controlled generation of coupled shear ($S_{xz}$) and uniaxial in-plane (IP) strains ($\delta S = S_{xx} - S_{yy}$) during lateral displacement between the stamp and substrate. These coordinated strain components synergistically address two critical challenges:

1. Bandgap engineering via OP strain-induced lattice deformation
2. Compensation of intrinsic structural asymmetries through IP strain-mediated symmetry restoration, effectively suppressing the FSS that arises from crystalline imperfections in QDs

This dual functionality establishes a robust platform for simultaneous spectral tuning and quantum state opti-mization in solid-state quantum emitters[32, 33].

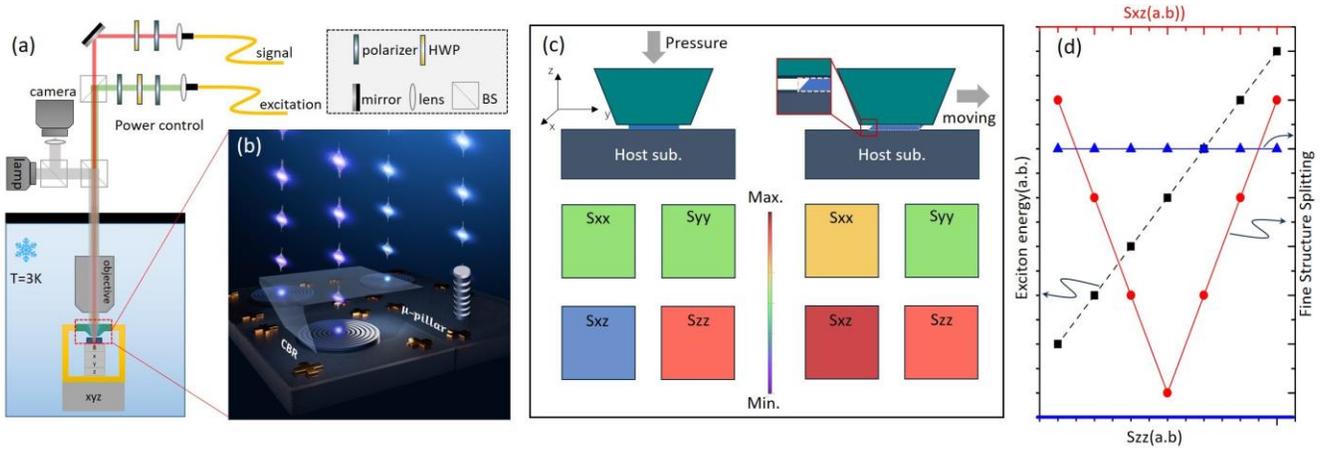

FIG. 2. **The sketch of the cryo-strain apparatus and the principle to generate 3D strain** (a) The optical setup of a helium bath cryostat. Both excitation laser and PL are fiber-coupled. HWP: half-wave plate; BS: beam splitter. (b) The strain apparatus consists of a micro-PDMS stamp and a multi-axial nano-positioner housed in a home-made metal box. The CBR sample is glued on a rotation stage fixed on a three-axial nano-positioner. (c) Pressing the micro-PDMS against the sample generates OP strain, as well as IP strain. The IP component is biaxial i.e. Sxx=Syy and is homogeneously distributed(left panel). Lateral movement of the micro-PDMS or the host substrate while maintaining the pressure introduces shear and uniaxial strain. The grey-dashed rectangle indicates the original shape the sample. Strain component Sxx is much larger than Syy(right panel). (d) The exciton energy and FSS as a function of strain. The exciton energy changes linearly with Szz stress and Sxz(black curve). The FSS features a 'V' shaped tuning behavior with Sxz stress (red curve), while remaining almost constant with Szz(blue curve).

**Principle of 3D strain generation**

The sample is an InAs/GaAs QD deterministically embedded in a CBR, as sketched in Fig. 2b. This cavity enables high extraction efficiency over a broad bandwidth. Further details regarding the device fabrication process can be found in Ref.[34]. Utilizing the finite element method (FEM), we conducted a comprehensive investigation into the strain profile generated when a compliant PDMS stamp is pressed against QDs, with simulated results depicted in Fig. 2c. The OP strain component $S_{zz}$ along the z-axis shows uniform distribution across the substrate contact area, while concurrent IP biaxial strains $S_{xx}$ and $S_{yy}$ ($S_{xx} = S_{yy}$) emerge due to PDMS expansion. The shear strain $S_{xz}$ is observed to be an order of magnitude smaller than these primary components, justifying its negligible influence in subsequent analyses. To disrupt in-plane symmetry, lateral displacement (e.g., along the x-axis) is introduced under sustained downward pressure. The viscoelastic properties of PDMS maintain interfacial cohesion with QDs, preventing slippage while enabling controlled strain transfer within practical displacement limits. This asymmetric loading protocol generates three dominant strain components: $S_{xx}$, $S_{yy}$, and $S_{xz}$. Notably, $S_{xx}$ becomes markedly larger than $S_{yy}$ under these conditions, contrasting with their initial equivalence in the symmetric case. We define the differential strain $\delta S = S_{xx} - S_{yy}$ as the *effective uniaxial strain*, while the shear component $S_{xz}$ gains significance compared to its initially trivial magnitude. Other shear strains (e.g., $S_{yz}$) remain negligible, being at least an order of magnitude smaller than dominant terms.

Figure 2d demonstrates strain-dependent modulation of exciton energy ($X$) and FSS, where $X$ exhibits linear correlation with applied strain. OP strain $S_{zz}$ along the QD growth direction minimally affects FSS (blue curve) due to preserved symmetry, whereas shear/IP strains ($S_{xz}/\delta S$) induce pronounced V-shaped FSS tuning (red curve) through symmetry modification. The theoretical framework for this strain-FSS relationship is detailed in the Methods section, followed by experimental validation of strain-mediated exciton engineering strategies.

**Vertical strain tuning of QDs**

The QDs in CBR are excited via above-band excitation. The image of a QD-coupled CBR cavity is depicted in extended Fig. 1(a,b,c). Previous attempts of strain tuning of the CBR utilize in-plane strain generated by a piezo substrate[30]. However, the central QD within a CBR is insulated from external materials, making it challenging to transfer in-plane stress to the QDs. This isolation has resulted in a remarkably restricted tuning range, achieving no more than 1 meV. This limitation can be overcome by implementing an OP strain application. Unlike IP strain, the transfer of OP strain remains unimpeded by the dimensions and configuration of the sample.

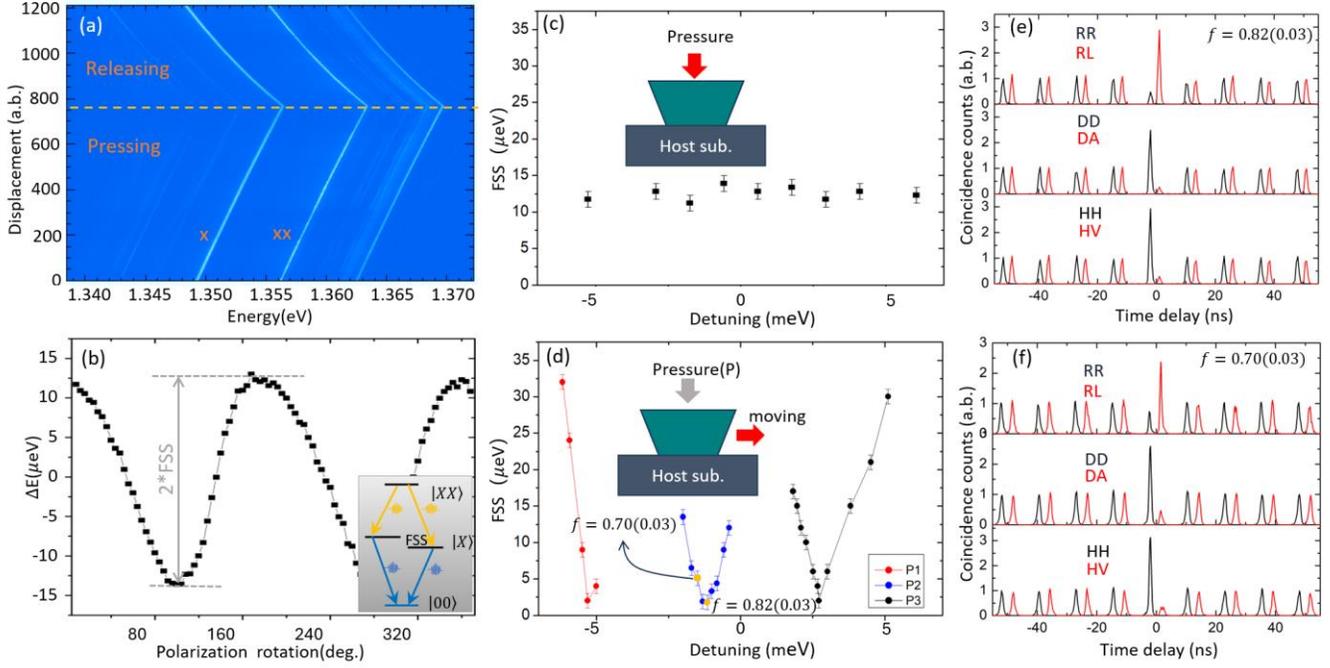

FIG. 3. **FSS tuning with IP strain and OP strain.** (a)The spectra of tuning QDs with vertical strain, exhibiting large and reversible tuning of X and XX. (b) FSS measurement. Polarization dependence of the energy difference between XX and X, showing a sinusoidal function. **Inset:** The XX-X cascade diagram. The FSS is defined as the energy difference between two exciton states. (c) The FSS tuning under OP strain. The exciton energy is shifted more than 10 meV while the FSS remains almost constant. (d) The FSS tuning with shear and IP uniaxial strain. The X energy and the FSS can be independently controlled.(e) Cross correlation reveals a entanglement fidelity of 0.81(0.03) for a near-zero FSS (∼ 2μeV) QD. (f) The entanglement fidelity drops to 0.70(0.03) when the FSS is tuned to ∼ 5μeV.

The PDMS stamp, with its low Young's modulus at cryogenic temperatures, serves as an exemplary medium for strain transfer. Upon upward actuation of the sample, the PDMS stamp exerts compression, thereby introducing OP strain within the QD and causing a blue-shift in its emission peak. Conversely, the release of this strain results in a corresponding red-shift, as depicted in Fig. 3a, which showcases the reversible tunability of the X emission energy. To the best of our knowledge, this is the first demonstration of applying OP strain to QDs. Compared to other tuning techniques, this method offers a large tuning range while upholding high precision. We have reliably produced over 700 distinct spectra within a 5 meV span, achieving a resolution of 7 μeV, a finesse comparable to the intrinsic full width at half maximum of the spectrum. The X emission is very sensitive to the OP strain, the tuning range can be further extended to tens of meV through the application of increased pressure.

The PDMS application barely induces degradation in photon collection efficiency of 23% while producing a characteristic ∼5nm redshift of cavity modes (see Supplementary Information, SI). This wavelength displacement arises from the modified local dielectric environment. For practical device implementation, a proactive design compensation can be incorporated through targeted cavity length adjustment to offset the PDMS-induced spectral shift while maintaining optimal collection efficiency.

While strain tuning demonstrates excellent temporal stability (SI), the asymmetric tuning behavior observed in Fig 3a reveals non-reproducible spectral shifts between compression and retraction cycles. Crucially, the non-repeatability presents no operational limitation as we implement real-time spectral verification to confirm target wavelength acquisition.

**FSS tuning with in-plane strain**

The XX energy level is illustrated in the inset of Fig. 3b. The FSS, a result of reduced in-plane symmetry, is defined as the energy difference between intermediate X states. For biaxial strain, the change in FSS caused by $S_{xx}$ is balanced out by $S_{yy}$, resulting in negligible effects on FSS tuning[35]. Therefore, when pressing the sample against the PDMS stamp, only the emission peaks are shifted and the FSS should remain unchanged. To verify this point, the emitted photons are directed to a spectrometer, with a half-wave plate and polarizer inserted into the beam path for polarization mapping. The FSS is extracted by fitting the energy difference between the XX and X peaks as a function of the polarization rotation angle with a sinusoidal function as shown in Fig. 3b. A typical mapping spectra are plotted in Extended Data Fig. 2a. The FSS is recorded as a function of wavelength, as plotted in Fig. 3c. Notably, the emission peaks exhibit a pronounced shift exceeding 10 meV, while the FSS maintains its stability, which confirms that OP strain shifts the emission peak without any significant modification to the FSS.

In addition to wavelength tuning, the precise control of the FSS is crucial for generating entangled photon pairs. It is important to note that QDs possess unique properties, including wavelengths, dipole orientations, and FSS[36]. The ability to erase the FSS depends on the direction of applied stress. In our experimental setup, the sample mount is designed with rotational freedom, enabling precise alignment of the strain with the dipole orientation of the QDs. For well-aligned QDs, with pressure maintained at a constant value P1, we systematically translate the QD along the x axis to exert $\delta$ S and Sxz, as illustrated in Fig. 3d. Once again, we record the FSS as a function of the X energy. FSS was suppressed from >30$\mu$eV to $\sim$ 2$\mu$eV , a value compatible with entangled photon generation but still above the radiative linewidth ($\sim$ 1$\mu$eV). The reason for non-vanishing FSS could be that the QD is not well aligned with the strain axis. Further reduction of residual FSS may require better alignment with multi-axial strain control [26]. This presents a stark contrast to the previous OP strain tuning. The FSS and X energy are strongly coupled, resulting in fixed wavelengths for entangled photons.

For the realization of scalable entangled photon sources, achieving independent tuning of the FSS and X energy is of enormous advantage. In the current case, we can utilize OP stress to tune the wavelength, while $\delta$S and Sxz to restore the FSS. When increasing the OP pressure to P2 or P3, the FSS tuning by $\delta$S and Sxz is plotted in Fig. 3d, indicated by the blue curve and black curve. At near-zero FSS, the energy of the exciton differs by approximately 7 meV compared to the previous case. Thus, we experimentally verified the independent control of FSS and wavelength. At near-zero FSS ($\sim$ 2$\mu$eV), quantum state tomography revealed a Bell-state fidelity of 0.82 ± 0.03 (Fig. 3e)[37]. This confirms that our strain-tuned source generates high-quality entangled photons for the realization of quantum repeaters. As comparison, we have performed quantum state tomography when the QD is tuned to a larger FSS ($\sim$ 5$\mu$eV). The corresponding entanglement fidelity drops to 0.70 ± 0.03 (Fig. 3f), suggesting the entanglement fidelity exhibits a strong dependence on FSS[38]. This underscores the critical role of FSS elimination in realizing practical quantum repeaters based on highly-entangled quantum light soures.

While our platform enables spectral tuning and FSS suppression, photon indistinguishability is another key requirement for quantum repeaters based on two-photon interference. The current device features an appreciable photon indistinguishability close to 0.7 (see SI), which can be further improved by cleaning the charge environment via electrical gating[39].

**Micropillar strain engineering**

The micropillar is considered one of the optimal cavity structures to generate high-quality single photons. However, the wide range tunability of QDs in micropillars has not been achieved to date. In such high aspect ratio nanostructures, strain engineering is typically considered unfeasible, as IP stress cannot be transferred to the QDs. Recently, a compromised solution by integrating shallowly etched pillars with a piezoelectric ceramic substrate is tested, where a tuning range of less than 1 meV is achieved[29]. In the following, we show that this challenge can be well addressed via the OP in-situ strain engineering.

The micropillars utilized in the experiment have a diameter of around 2.9 $\mu$m. The images of these pillars are shown in Extended Data Fig. 1(d,e,f). The mode profile is simulated using the finite time domain difference method. Since we employ fiber coupling to collect the PL in the experiment, only two modes that have solid electric fields in the center are plotted. The mode profiles are depicted in Fig. 4a. These modes i.e. the HE11 mode and the HE12 mode, are capable of well-coupling to a single-mode fiber. The HE11 represents the fundamental mode and HE12 the high-order mode[40]. Experimentally, as plotted in Fig. 4b, two modes emerge when the device is strongly pumped with 785 nm CW laser. The mode spectrum and mode profiles agree well with the simulations. As the PDMS stamp applies stress to the pillar, both the QD emission peak and cavity mode shift simultaneously but at different speeds. The cavity modes shift at a slower rate (see Extended Data Fig. 2b). The spectra illustrated in Fig. 4c are manually shifted to maintain the cavity mode constant so that the QD-cavity detuning can be clearly resolved.

Initially, the emission peak, which is detuned from the fundamental cavity mode HE11, appears weaker. As the emission peak and the mode approach resonance, a significant enhancement in PL is observed, indicating efficient light-matter coupling. To extract the Purcell factor, we have conducted time-resolved resonance fluorescence measurements for different QD-cavity detuning, as depicted in Fig. 4c. And the intensity of the X line while scanning the cavity mode is plotted as an inset, showing more than 14x enhancement in intensity. All time traces are fitted with mono-exponential decay pulses deconvoluted with instrument response function (IRF). In a far-detuned case, a time constant of 300 ps was recorded, represented by the blue curve in Fig. 4d. When the QD-cavity is tuned into resonance, the lifetime is shortened to 80 ps (time trace with the red curve in Fig. 4d). Inset is the lifetime of a QD embedded in bulk substrate which is $\sim$ 1$ns$. By comparison, a 12.5 fold radiative rate enhancement can be extracted. Fig. 4e shows the lifetime of X at various detuning. Detailed time traces are plotted in Extended Data Fig. 3. Furthermore, due to the large tuning range, we can even scan the emission peak across the high-order mode HE12, enabling the generation of single photons in a different spatial profile and polarization state.

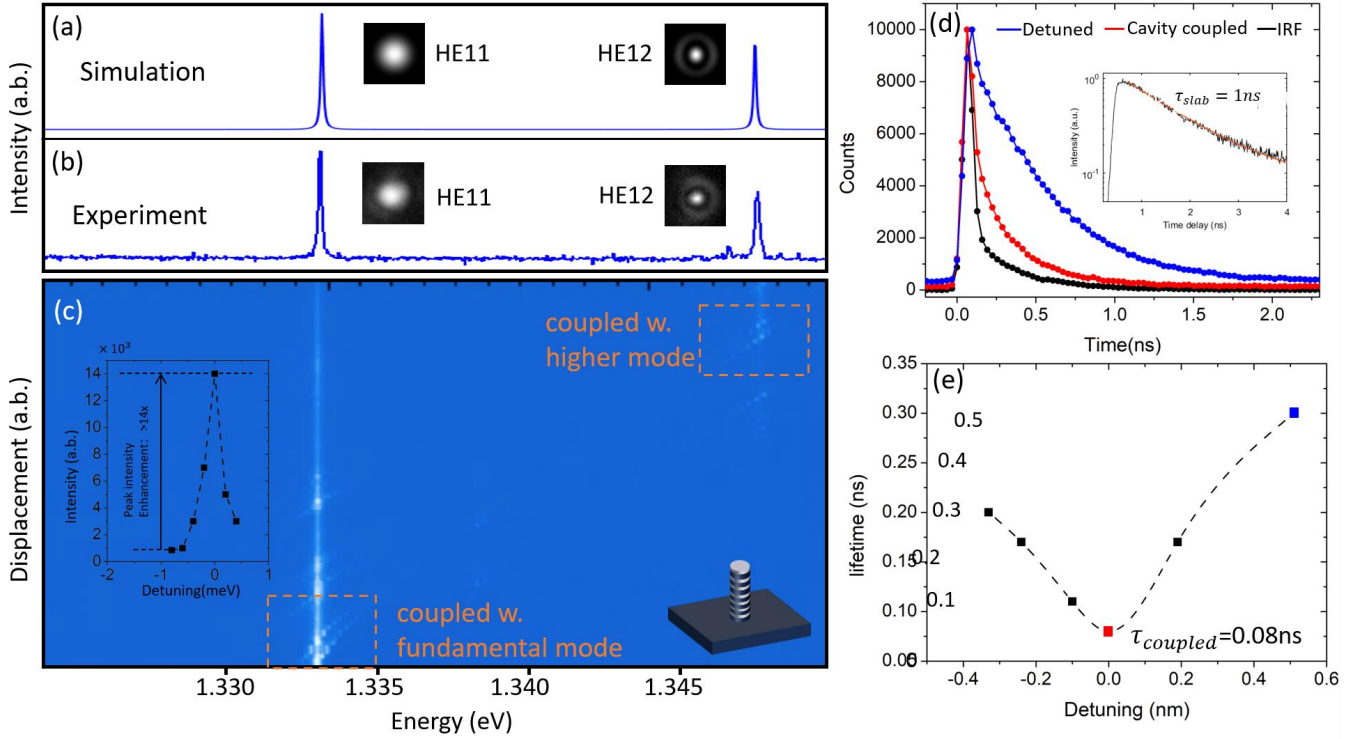

FIG. 4. **Micropillar strain engineering.** (**a**) The simulated and (**b**) measured cavity modes of the micropillar. (**c**) The spectral mapping of the QD's emission and cavity modes with strain. The QD's emission peak is scanned across both the fundamental mode and a high-order mode. At resonance, the emission intensity is significantly enhanced. (**d**) The lifetime traces of X for the detuned (blue line) and resonant cases (red line). Inset shows the lifetime of a QD embedded in bulk. (**e**) The lifetime of X as a function of QD-cavity detuning.

**Conclusion**

To conclude, we introduce an in-situ 3D strain at cryogenic temperatures, showcasing its effectiveness in independently controlling the FSS and exciton energy of single QDs embedded in CBRs. We also demonstrate the large-range tuning of QDs in micropillars, where the QD emission is tuned across both the fundamental mode and a high-order mode. Our work makes an important step towards the scalable quantum network in which the links and communications between remote quantum nodes are based on high-performance quantum light sources with highly engineerable optical characteristics. These results also highlight the versatility of micro-PDMS stamps, which can be employed e.g., for in-situ manipulation of van der Waals heterostructures at low temperatures[41, 42]. Additionally, the compatibility of 3D strain tuning with other techniques, such as electric-field tuning and magnetic-field tuning, offers significant advantages. Compared with other cryo-strain apparatus, our device is advantageous in terms of magnitude, strain form (3D) and in-situ operation, as is discussed in Extended Data table 1 and will also find extensive interest in the broader community of condensed matter physics.

---

* chenyanxyz@outlook.com
† equal contribution
‡ tjiang@nudt.edu.cn
§ liujin23@mail.sysu.edu.cn

**Acknowledgements** This research is supported by the National Natural Science Foundation of China (12374476, 62035017, 12334017, 12293052, 12104522), the Natural Science Foundation of Guang-dong (2022A1515011400), Guangdong Introducing Innovative and Entrepreneurial Teams of "The Pearl River Talent Recruitment Program" (2021ZT09X044), The innovative research program of National University of Defense Technology (22-ZZCX-067). Guang-dong Basic and Applied Basic Research Foundation (2023B1515120070).

**Author contributions** T. J, Y. C and J. L conceived the project. Y. C performed the numerical simulations. Y. C, X. L and S. L fabricated the devices. X. L, S. L, J. Y, Y. W P. Y and K. X built the setup and characterized the devices. P. C., X. L, C. Z, Y. Y analyzed the data. Y. C and J. L wrote the manuscript with inputs from all authors.

**Competing interests** The authors declare no competing financial interests.

**Data and materials availability** The data sets will be available upon reasonable request.